\documentclass[prc,superscriptaddress,unsortedaddress,twocolumn]{revtex4}
\usepackage{graphicx,epsf,amsmath,color}
\usepackage{ulem}
\def\bp{\mbox{\boldmath $p$}}
\def\bq{\mbox{\boldmath $q$}}
\def\br{\mbox{\boldmath $r$}}
\def\bR{\mbox{\boldmath $R$}}

\begin{document}
\title{$\Xi$-deuteron low-energy $s$-wave phase shifts and momentum correlation functions
in Faddeev formulation}
\author{M. Kohno}
\affiliation{Research Center for Nuclear Physics, Osaka University, Ibaraki 567-0047,
Japan\\}

\author{H. Kamada}
\affiliation{Research Center for Nuclear Physics, Osaka University, Ibaraki 567-0047, Japan\\}
\affiliation{Department of Physics, Kyushu Institute of Technology, 1-1 Sensuicho, Tobata,
Kitakyushu 804-8550, Japan\\}
\begin{abstract}
The low-energy $\Xi$-deuteron scattering is investigated through the solution of Faddeev equations,
employing three sets of the currently available representation of the $\Xi$-nucleon interactions.
One of these is the chiral NLO interaction specified by the J\"{u}lich group,
and the other two
are based on the calculations by the HAL-QCD method. The $s$-wave phase shifts in the
$J=3/2$ and $J=1/2$ states are presented. Three-body wave functions in coordinate space
are constructed from the Faddeev amplitudes in momentum space. These functions are used in
the calculation of $\Xi d$ momentum correlation functions. The effects of the deuteron breakup
are significant in the $J=3/2$ channel. The differences in the magnitude of the calculated
correlation function show the quantitative difference of the $\Xi N$ interactions in the
spin-isospin channels. The prospective experimental data on the $\Xi d$ momentum
correlation function could contribute to a better description of the $\Xi N$ interactions. 
\end{abstract}
\maketitle

\section{Introduction}
It is essential to understand the interactions between hyperons and nucleons in order
to explore the role of the strangeness in the hadron world, contingent on its presence
or absence. However, the direct hyperon-nucleon ($YN$) scattering experiments
are challenging at present, though some progresses have been reported \cite{SIGP22}.
In the meantime,
the hyperon-nucleon momentum correlation functions measured in heavy-ion
collision experiments have been identified as a promising source of information
of the relevant interaction. The $\Lambda$-proton and $\Xi$-proton correlation
functions \cite{ACHA19,ALICE22} are used to study and improve the theoretical description of the
$YN$ interactions. The experimental efforts to measure the hyperon-deuteron ($Yd$)
correlation functions are progressing \cite{STAR25}, providing further valuable information
about the $YN$ interactions. It is noted that because the correlation functions
measured in heavy-ion collision experiments do not directly arise from a genuine two-body
process and the theoretical description of them depends on the assumption represented
by the source function, the off-shell ambiguity of the underlying baryon-baryon
interactions has to be kept in mind as pointed out in Ref. \cite{EPEL26}.
 
We have performed Faddeev calculations of the $\Lambda d$ low-energy scattering
and calculated the $\Lambda d$ correlation function with the Faddeev three-body
wave function in the scattering state \cite{KK24,KK25}. In the present article,
the method of the Faddeev calculation is extended to the $\Xi d$ correlation function,
using three sets of the $\Xi N$ interaction that are recently developed.

Several $\Xi$ bound states in light nuclei observed recently in emulsion experiments
\cite{NAK15,HAY21,YO21,IC24} indicate that the $\Xi N$ interaction is attractive
on the average to support the $\Xi$ bound states, although the experimental
uncertainties are fairly large. These experimental data
are valuable, but it is not straightforward to figure out the spin-isospin structure
of the $\Xi N$ interaction from the information on the $\Xi$ bound-state energies,
because four spin-isospin channels are possible for every partial wave due to
the absence of the Pauli exclusion on the $\Xi N$ two-body system.

On the theoretical side, the construction of baryon-baryon interactions in the
strangeness $S=-2$ sector has been progressing based on the fundamental theory of
hadrons and their interactions, within the framework of chiral effective field theory (ChEFT)
or through the use of lattice calculation methods for the quantum chromodynamics (QCD).
We employ the following three descriptions of the $\Xi N$ interaction.
One is the description in the next-to-leading order (NLO) in ChEFT, which was specified
by the J\"{u}lich-Bonn-M\"{u}nchen group \cite{HAID16,HAID19} based on limited
experimental data. Others are parametrizations by Inoue {\it et al.} \cite{Ino19} and
by Sasaki {\it et al.} \cite{Sas20} based on the HAL-QCD method.

The present manuscript focuses on the deuteron breakup contributions and their
dependence on the $\Xi N$ interactions, which can be demonstrated by ignoring the
Coulomb interaction. In instances where the incorporation of the Coulomb interaction
is essential for the analysis of experimental data, a method for including the Coulomb effect
can be employed. In the context of proton-deuteron scattering, an accurate
method for incorporating the Coulomb force has been recently developed \cite{Del02,Wit24}.
The Faddeev calculations are
carried out in an isospin basis with the averaged masses for nucleons and hyperons;
938.919 MeV, 1115.68 MeV, 1193.12 MeV, and 1318.29 MeV for $N$, $\Lambda$,
$\Sigma$, and $\Xi$, respectively. Therefore, the present calculations may not be realistic
enough. Nevertheless, otherwise fully microscopic Faddeev calculations for the $\Xi d$
scattering states and correlation functions can provide basic information how the
three-body dynamics influence the $\Xi d$ system, which is useful in studying the
properties of the existing $\Xi N$ interactions in view of the prospective experimental data. 

The effect of the deuteron breakup on the $\Xi d$ correlation function has been examined
by Ogata \textit{et al.} \cite{Og21}, employing the folding procedure for the $\Xi$-d potential
with the AV4' NN \cite{AV4D} and the HAL-QCD $\Xi N$ interaction
of Sasaki \textit{et al.} \cite{Sas20}. They reported that the breakup effect is appreciable
but not very significant. In the present calculations, the Faddeev formulation is employed
to rigorously treat a larger number of three-body correlations. Consequently, significant
effects may emerge arising from rearrangement channels.  

In Sec. 2, the basic properties of the three $\Xi N$ interactions employed are surveyed
by showing their $\Xi N$ phase shifts. $\Xi$-deuteron elastic phase shifts obtained by
the Faddeev calculation are presented in Sec. III.  In Sec. IV,  the theoretical expression
of the $\Xi d$ momentum correlation function is recapitulated. The construction of the
$\Xi d$ three-body wave function from the Faddeev amplitudes including
the rearrangement channel is also reviewed. Calculated results the $\Xi d$ 
correlation functions for the three $\Xi N$ interactions are presented in Sec. V.
Besides the results with the full three-body wave function, those with the incident-channel
three-body wave function are also shown, although they cannot be separately observed
in experiments. Summary follows in Sec. VI.

\section{$\Xi N$ phase shift}
\begin{figure}[t]
\centering
 \includegraphics[width=0.45\textwidth,pagebox=cropbox,clip]{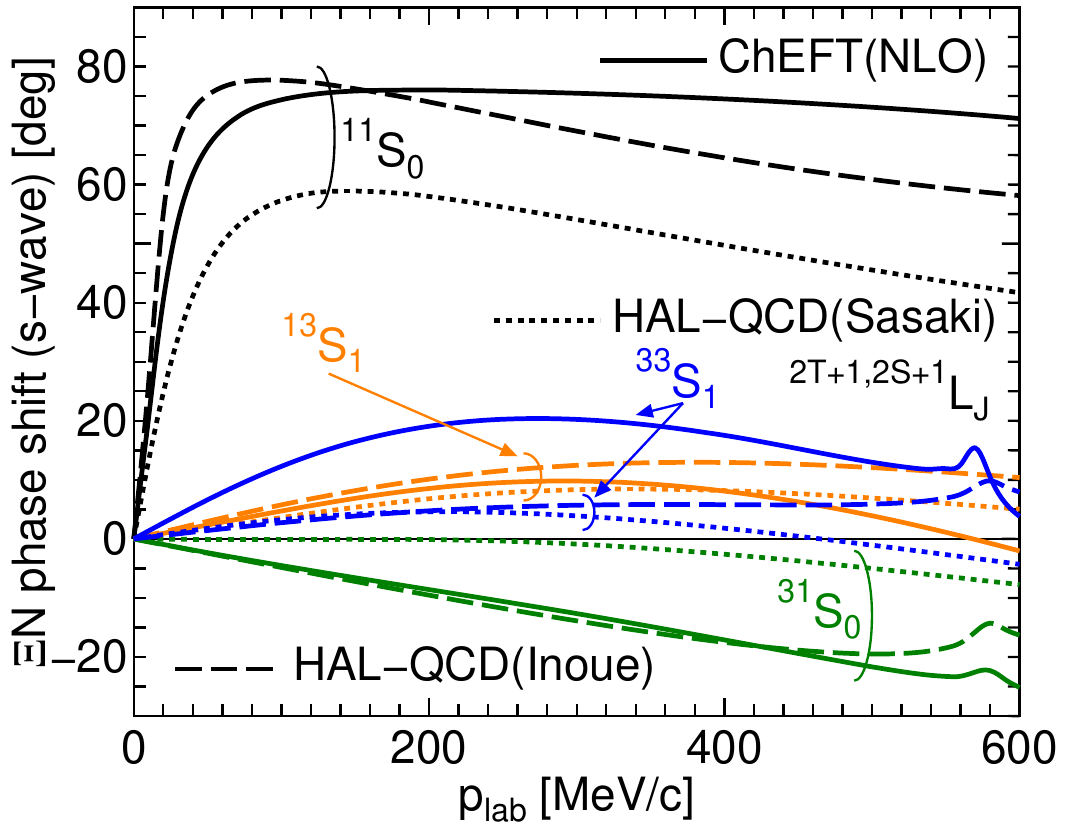}
\caption{
$\Xi N$ $s$-wave phase shifts calculated with NLO ChEFT interactions
($\Lambda_c=550$ MeV) \cite{HAID19} are shown by solid curves
with the notation $^{2T+1,2S+1}L_J$ for specifying the spin $S$ and isospin $T$
channel. Phase shifts
with two sets of the parametrization based on the HAL-QCD calculations are
also shown: one is with the full parametrization by Inoue \textit{et al.} \cite{Ino19} (dashed)
and the other is with the fitting parameters for $t/a=12$ by Sasaki \textit{et al.} \cite{Sas20} (dotted).
}
\label{fig:xnsph}
\end{figure}

It is requisite before the Faddeev calculation of the $\Xi d$ system to survey the
properties of the relevant two-body $\Xi N$ interactions employed in this article
by evaluating the phase shifts of $\Xi N$ elastic scattering. The $s$-wave phase
shifts in each spin-isospin channel with three representations of
the $\Xi N$ interaction were reported in Fig. 1 of Ref. \cite{KM21} and cited here
as Fig. \ref{fig:xnsph}. The spin(S)-isospin(T) channel with the total
angular momentum $J$ is denoted as $^{2T+1,2S+1}S_J$.
The solid curves represent the results of an updated version of the chiral NLO $S=-2$
interactions \cite{HAID19} with the cutoff parameter of $\Lambda_c=550$ MeV
that are formulated in momentum space. Other curves represent the phase shifts with
the interactions parametrized as a local potential on the basis of HAL-QCD calculations.
The dashed curves are the results of the potential by Inoue
{\textit{et al.}} \cite{Ino19}. The dotted curves are the results of the potential with
the fitting parameters for $t/a=12$ by Sasaki \textit{et al.} \cite{Sas20} in which the effects
of the tensor coupling and the baryon-channel coupling except for
$\Lambda\Lambda$ are renormalized to a local $\Xi N$ central potential.
Consequently, this interaction is not suitable for considering in-medium effects
to the $\Xi N$ interaction. These two HAL-QCD potentials are referred
as the Inoue potential and the Sasaki potential, respectively, hereafter.

Three potentials appear to predict qualitatively similar behavior of the $\Xi N$
phase shifts. That is, the interaction in the $^{31}S_0$ state is repulsive,
and the interactions in the remaining three states are attractive. Nevertheless,
some quantitative differences are remarked. The repulsive character in the
$^{31}S_0$ part of the Sasaki potential is very weak.
The attraction in the $^{33}S_1$ state of the HAL-QCD parametrization is
weaker than that of ChEFT. The attraction in the $^{13}S_1$ state, in which
no baryon-channel coupling is present,  is weak.
The $^{11}S_0$ state is strongly attractive, although no bound state is expected.
The couplings to the $\Lambda\Lambda$ as well as $\Sigma\Sigma$ states are
important for this attractive property. The attractive strength in the $^{33}S_1$ state
is not so large as in the $T=0$ $^{11}S_0$ state, but plays an important role
in many body-systems due to the spin-isospin weight factor of $(2S+1)(2T+1)=9$.

The calculated results of the $\Xi d$ correlation function presented in the following reveal
the consequence of these quantitative differences to the momentum correlation function
which is determined by the three-body wave function in the area within the source radius. 

\section{$\Xi$-deuteron $s$-wave phase shifts}
In this section, the $\Xi d$ $s$-wave elastic scattering phase shifts calculated in a Fadeev
formulation are presented using three $YN$ interactions that were explained in the previous
sections. The structure of the Faddeev equations is the same as those for the $\Lambda d$
scattering except for the inclusion of the isospin degrees of freedom of the $\Xi$ hyperon.

\begin{figure}[t]
\centering
 \includegraphics[width=0.45\textwidth,pagebox=cropbox,clip]{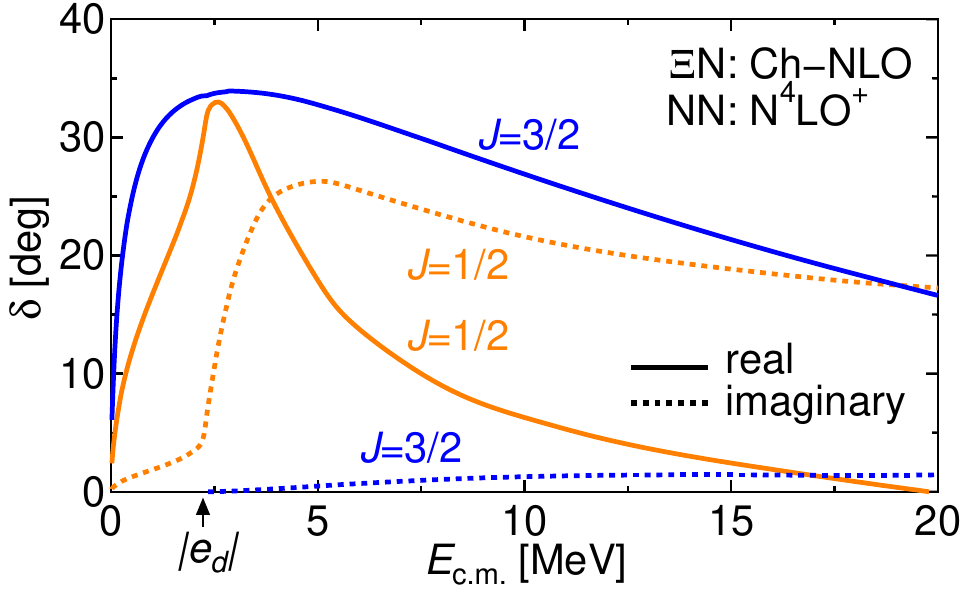}
 \hspace{0.3em}
 \includegraphics[width=0.45\textwidth,pagebox=cropbox,clip]{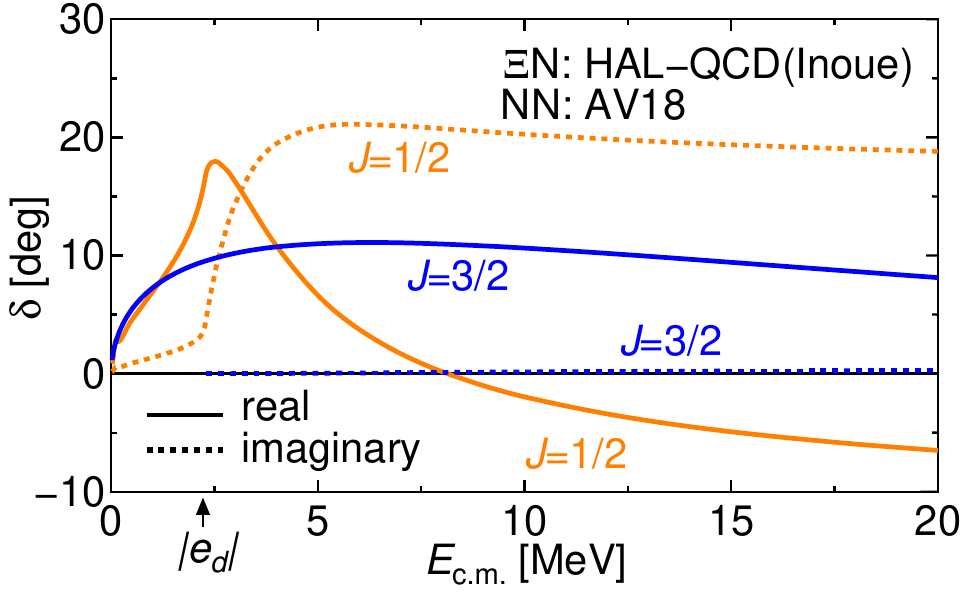}
 \hspace{0.3em}
 \includegraphics[width=0.45\textwidth,pagebox=cropbox,clip]{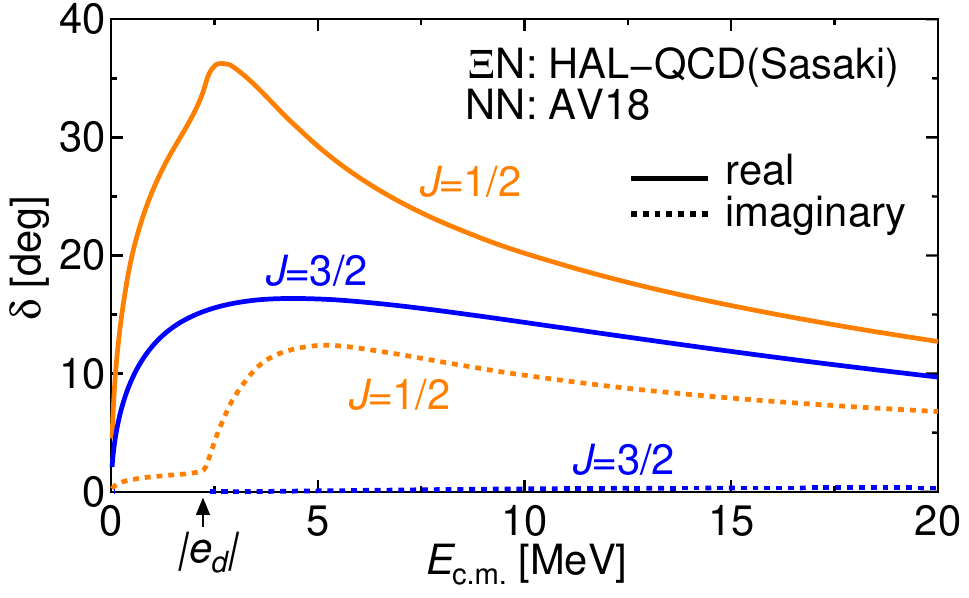}
\caption{
$\Xi d$ $s$-wave phase shifts in the $J=1/2$ and $J=3/2$ channels calculated in
the Faddeev method. the upper panel shows the results with the chiral
NLO $S=-2$ interaction \cite{HAID19} with the chiral N$^4$LO$^+$ $NN$
interaction \cite{RKE18}. The middle and lower panels depict the results with
the HAL-QCD Inoue and Sasaki potentials \cite{Ino19, Sas20}, respectively.
For the HAL=QCD interactions, the AV18 interaction \cite{AV18} is used for the $NN$ part.}
\label{fig:xdph}
\end{figure}

Fig. \ref{fig:xdph} presents the results for the total angular momentum $J=1/2$ and $3/2$
channels, employing the three $\Xi N$ interactions: the chiral NLO, Inoue and Sasaki potentials.
The low-energy behavior of the phase shifts indicates that no $\Xi NN$ bound state is
expected in both $J=1/2$ and $3/2$. The absence of the bound state with the chiral
NLO interaction is consistent with the result of searching the $\Xi NN$ bound state by
Faddeev calculations reported in Ref. \cite{MK21}.

In the $J=3/2$ $\Xi d$ state, the spin singlet $\Xi N$ interactions are not involved.
The difference of the magnitude of the phase shift among three
potentials reflects the difference of the strength of the attraction in the $^{13}S_1$ and
$^{33}S_1$ $\Xi N$ interactions. The imaginary part of the phase shift is zero below
the deuteron threshold, $E_{cm}=|e_d|$, because there is no transition to an energetically
possible $\Lambda \Lambda N$ state because the spin triplet $\Lambda\Lambda$ $s$
state is not possible due to the Pauli principle. While the imaginary phase shift is not zero
above the deuteron breakup threshold, its magnitude is small because
the deuteron breakup is not possible into the isospin singlet $np$ channel.      

In the $J=1/2$ channel, all four spin-isospin states participate in the $\Xi d$ interaction.
The peak structure around the deuteron threshold is due to the coupling to the $^1S_0$ $NN$
state due to the strong attraction realized as a virtual pole structure. The large imaginary
phase shifts in contrast to those in the $J=3/2$ case are due to the $^1S_0$ $NN$
open state above the deuteron threshold. In the $J=1/2$ case, the $\Xi d$ system can
decay to the $\Lambda\Lambda N$ state below the deuteron threshold. However, the
magnitude of the imaginary phase shift is small because the $\Xi N$-$\Lambda\Lambda$
coupling is weak. The relatively large $J=1/2$ phase shifts of the Sasaki potential is caused
by the minimal repulsive $^{31}$S$_0$ interaction compared with other potentials, as is seen
in Fig. \ref{fig:xnsph}. 

The scattering length $a_s$ and the effective range $r_e$,
which are derived from the scattering amplitudes between $E_{cm}=0.02 - 1.6$ MeV, are presented
in Table I. They are complex in the $J=1/2$ channel. The magnitude of these parameters is
comparable to the parameters reported by Ogata \textit{et al.} in Ref. \cite{Og21}.

\begin{table}
 \begin{tabular}{c|cc|cc}\hline \hline
    & \multicolumn{2}{c|}{$J=1/2$} &  \multicolumn{2}{c}{$J=3/2$} \\
 $\Xi N$ interaction  &  $a_s$ & $r_e$ & $a_s$ & $r_e$ \\ \hline
  ChEFT & $-1.49-i0.15$ &  \phantom{$-$}$1.11-i0.54$ & $-3.79$ & $4.11$ \\
  Inoue &  $-0.53-i0.10$  & $-9.56+i1.69$  &  $-0.72$ & $8.65$ \\
  Sasaki &  $-2.80-i0.16$ & \phantom{$-$}$3.62-i0.16$  &  $-1.30$ & $7.00$  \\ \hline \hline
 \end{tabular}
\caption{The scattering length $a_s$ and the effective range $r_e$ of
$\Xi d$ scattering. Entries are in the unit of fm.}
\end{table}

\section{$\Xi$-deuteron correlation function}
The $\Xi d$ three-body wave functions including the breakup effects in both an incident
channel and a rearrangement channel are derived from calculated Faddeev amplitudes,
which are employed to evaluate $\Xi d$ momentum correlation functions. The theoretical
model of the correlation function and the procedure of obtaining the three-body wave function
were explained in Ref. \cite{KK25} for the $\Lambda d$ case. These expressions are summarized
in this section for the sake of completeness.

The derivation of the theoretical model description of the $\Xi d$ momentum correlation function
$C(q_0)$ starts from the following definition based on the formulation
by Mr\'{o}wczy\'{n}ski \cite{MR20}:
\begin{align}
 C(q_0)A_2=& g_sg_I (2\pi)^3 \int d \br_\Xi d \br_n d \br_p\; D(r_n)D(r_p)D(r_\Xi)
 \notag \\
 &\times |\psi(\br_\Xi,\br_n,\br_p;q_0)|^2.
\label{eq:def}
\end{align}
$\psi(\br_\Xi,\br_n,\br_p;q_0)$ is a $\Xi NN$ three-body wave function, the asymptotic
form of which is a free deuteron wave function $\phi_d(\br)$ times a $\Xi$-deuteron
relative plane wave with $q_0$ being the asymptotic relative momentum.
$g_s g_I=3/4$ is a statistical spin-isospin factor for the deuteron formation, and $D(r)$
is a Gaussian source function with the range parameter $R_s$ that represents the
distribution of pertinent baryons:
\begin{equation}
 D(r)=D(r;R_s)\equiv (2\pi R_s^2)^{-3/2}\exp\{-r^2/(2R_s^2)\},
\end{equation}
Although the source radius can be different between the nucleon and the $\Xi$ hyperon,
the same radius $R_s$ is assumed in the present article.
$A_2$ is a deuteron formation rate given by
\begin{equation}
 A_2= g_sg_I (2\pi)^3 \int d \br D_{np}(r) |\phi_d(\br)|^2,
\label{eq:a2}
\end{equation}
where the source function $D_{np}(r)$
for the deuteron is obtained by the convolution of the source function $D(r)$ as
\begin{align}
 D_{np}(r)=\int d \bR D(\bR+\frac{1}{2}\br)D(\bR-\frac{1}{2}\br)
 =D(r;\sqrt{2}R_s).
\label{eq:sfn}
\end{align}
Integrating out the center of mass coordinate and introducing a free $\Xi d$
wave function
\begin{equation}
\psi_0(\br_{np},\br_{3})=\phi_d(\br_{np})e^{i\bq_0\cdot \br_3},
\end{equation}
$C(q_0)$ in Eq. (1) is rewritten as
\begin{align}
 C(q_0)=&  1+\frac{\int d \br_{3} d \br_{np}\; D_{\Xi d}(r_{3})D_{np}(r_{np}) \Delta \psi}
{\int d \br D_{np}(r_{np})|\phi_d(\br)|^2},\\
 \Delta \psi \equiv& |\psi(\br_{np},\br_{3})|^2-|\psi_0(\br_{np},\br_{3})|^2.
\label{eq:cor1}
\end{align}
$\psi(\br_{np},\br_{3})$ is the $\Xi NN$ three-body wave function represented in the Jacobi
coordinates after removing the center-of-mass part from $\psi(\br_\Xi,\br_n,\br_p;q_0)$:
$\br_{np}$ is the relative coordinate between $n$ and $p$, and $\br_{3}$ is
the relative coordinate between $\Xi$ and the center-of-mass of the $np$ two-body system.
$D_{np}(r_{np})$ and $D_{\Xi d}(r_{3})$ are the source functions given by 
$D_{np}(r_{np})=D(r_{np};\sqrt{2}R_s)$ and $D_{\Xi d}(r_{3})=D_{\Xi d}(r_{3};\sqrt{2}R_s)$.
It is noteworthy that the source radius is different in $D_{np}$ and $D_{\Xi d}$.

It is reasonable in low-energy scattering to assume that only the $\Xi d$ $s$ wave deviates
from the plane wave. It may also be argued that the component of higher partial waves is
negligible in the short-range area imposed by the source function.
Then, $\psi(\br_{np},\br_{3})$ and $\psi_0(\br_{np},\br_{3})$
are to be replaced as
\begin{align}
\psi_0(\br_{np},\br_{3}) \rightarrow & \phi_d(\br_{np}) j_0(q_0r_{3}), \\
\psi(\br_{np},\br_{3}) \rightarrow & \sum_{\ell=0,2} (2\ell+1)i^\ell
 P_\ell(\cos\widehat{\bp \br_{np}}) \notag 
\\
 & \times \phi_\ell(p_{q_0}r_{np}) \varphi_0(r_{3};q_0),
\label{eq:cor2}
\end{align}
where $j_0$ is a spherical Bessel function and $\phi_\ell$ ($\varphi_0$) is a partial wave
in the coordinate $r_{np}$ ($r_3$).

In the case where the deuteron is supposed to be an elementary particle and the $\Xi d$
relative wave function $\varphi_0(r_{3};q_0)$ is represented by its asymptotic form
described by the effective range parameters, the expression of $C(q_0)$ is lead to
the Lednicky and Lyuboshits formula \cite{LL81}.

We evaluate the $\Xi d$ momentum correlation functions based on the expression given
by Eqs. (6) - (10), using the $\Xi d$ three-body wave function generated by the
Faddeev calculations.
The method of obtaining the three-body wave function from the Faddeev amplitudes
in momentum space was explained in Ref. \cite{KK25}. We can consider two types of
the $\Xi d$ three-body wave function. One is the wave function in an incident channel
which consists of an elastic part including a modification of the deuteron in an
interaction region and a remaining part representing a deuteron breakup. The other
is a full three-body wave function including effects of the deuteron breakup also
in a rearrangement channel. While the deuteron breakup effect in the incident
channel does not separately align with the experimental measurement,
it is nevertheless worthwhile to theoretically investigate its contribution to
understand the properties and the role of the $\Xi N$ interactions.

For reference, the basic equations employed for calculating the $\Xi d$ three-body
wave functions are explained in Appendix.
\section{Calculated results}
We present $\Xi d$ momentum correlation functions calculated by the $\Xi d$
three-body wave functions generated from the Faddeev amplitudes in momentum space,
using three $S=-2$ potentials explained in Sec. 2. First, the deuteron breakup effects
in the incident channel are shown. Although the correlation function by the incident
channel wave function is not directly related to the experimental data, it is theoretically
interesting to signifies the effect of the breakup in the incident channel.
In the previous calculations of the $\Lambda d$
correlation functions in Ref. \cite{KK25}, the transition to the $^1$S$_0$ $NN$
state from the deuteron state is not permitted and the effects of the deuteron
breakup contributions are found to be minimal. In the present case,
the $^1$S$_0$ $NN$ state can participate
in the $J=1/2$ channel. Because the $^1$S$_0$ $NN$ interaction is strongly
attractive which is manifested as a virtual pole structure, the coupling is expected
to affect the $\Xi d$ wave function to a large extent. Afterwords the results of the
full three-body wave function are presented. 

\subsection{Calculations with wave function in incident channel}
\begin{figure}[t]
\centering
 \includegraphics[width=0.45\textwidth,pagebox=cropbox,clip]{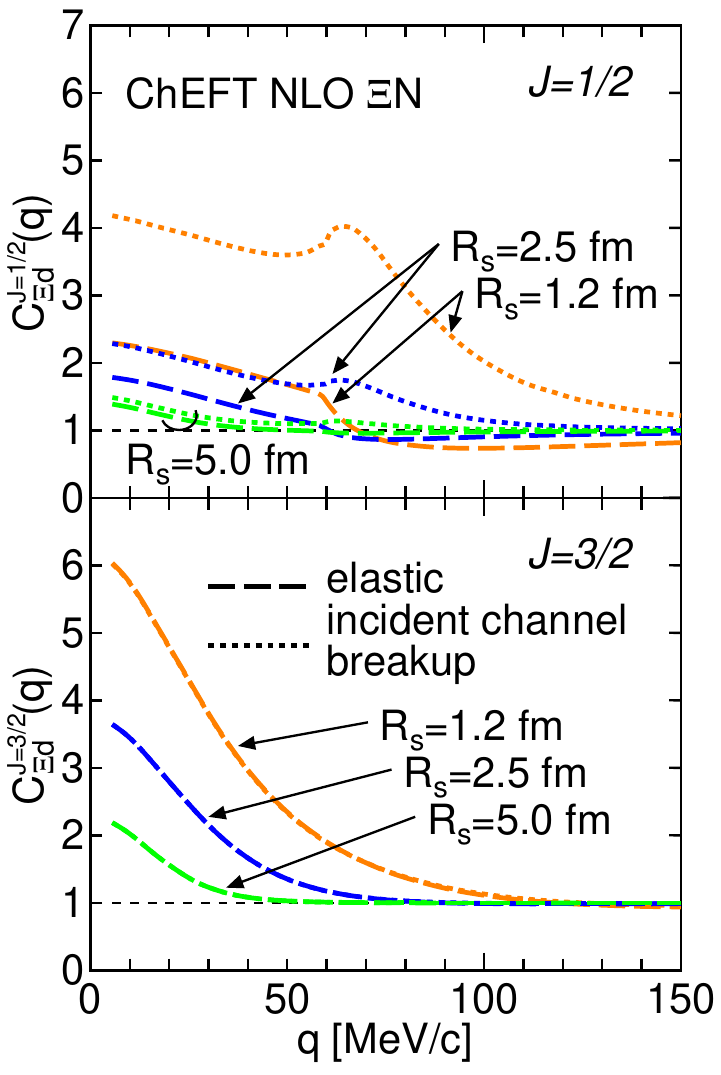}
\caption{$\Xi d$ momentum correlation functions in the $J=1/2$ and $3/2$ channels
for three choices of the source radius $R_s=1.2$, $2.5$, and $5.0$ fm, which are calculated
with the wave functions generated by the Faddeev calculations using the chiral NLO
$S=-2$ \cite{HAID19} and the chiral N$^4$LO$^+$ $NN$ \cite{RKE18} interactions.
The dashed curves are the results of the elastic wave functions.
The full incident channel wave functions yield the dotted curves. In the lower $J=3/2$ panel,
the dotted curves are not indistinguishable from the dashed curves. 
}
\label{fig:incch}
\end{figure}
Fig. \ref{fig:incch} shows the results of the chiral NLO interactions \cite{HAID19} together
with the chiral N$^4$LO$^+$ $NN$ interaction \cite{RKE18}, using the incident
channel $\Xi d$ wave functions. It should be noted that $C(q_0)$ is simply denoted by $C(q)$
in the subsequent figures. The cutoff scale is 550 MeV for both the $\Xi N$
and $NN$ interactions. Three representative source radius are selected for the calculations;
$R_s=1.2$, $2.5$, and $5.0$ fm, respectively.
The upper and lower panels for $J=1/2$ and $J=3/2$ are depicted
on the same vertical scale for the purpose of comparing their respective magnitudes.
The dashed curves are the results when the elastic $\Xi d$ wave functions are employed.
The dotted curves illustrate the results calculated by the wave functions of Eq. (A3) that
include the deuteron breakup in the incident channel. In the case of $J=3/2$, the
$^1$S$_0$ $NN$ state is not involved in the present restricted space. Therefore, the effect of the
incident-channel deuteron breakup is very small as in the case of the $\Lambda d$
correlation function \cite{KK25}. The dotted curves in the lower panel of Fig. \ref{fig:incch}
are indistinguishable from the dashed ones.

\begin{figure}[b]
\centering
 \includegraphics[width=0.45\textwidth,pagebox=cropbox,clip]{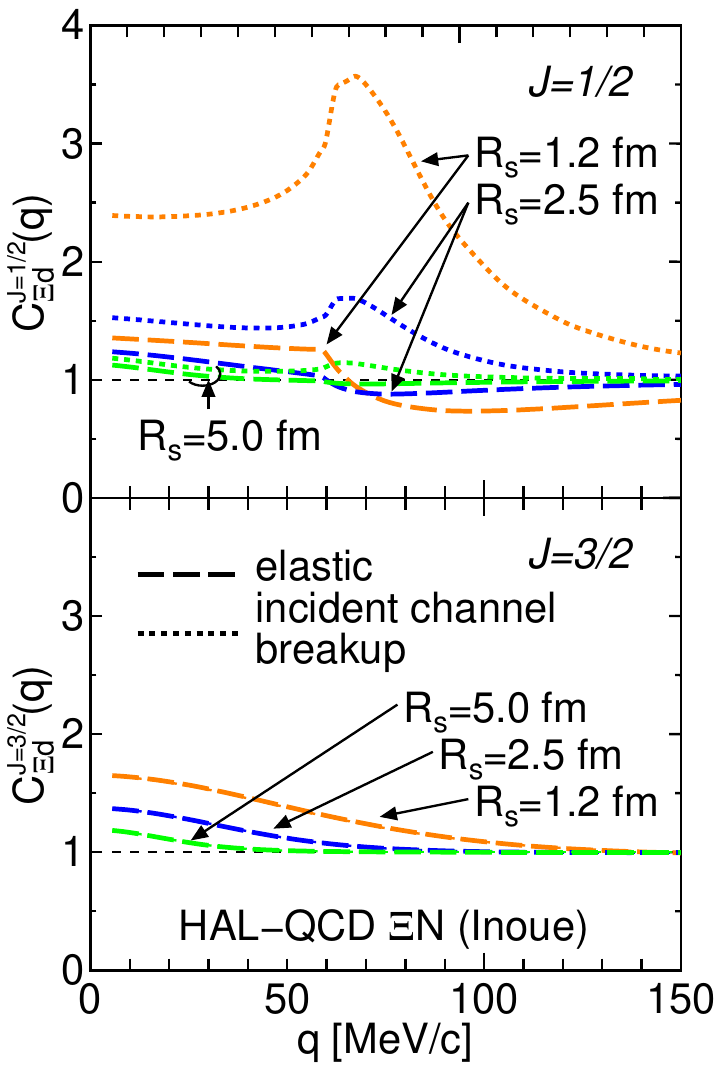}
\caption{Same as Fig. \ref{fig:incch}, but for the Inoue $S=-2$ potential \cite{Ino19} and
the AV18 \cite{AV18} $NN$ interaction.
}
\label{fig:incino}
\end{figure}

\begin{figure}[b]
\centering
 \includegraphics[width=0.45\textwidth,pagebox=cropbox,clip]{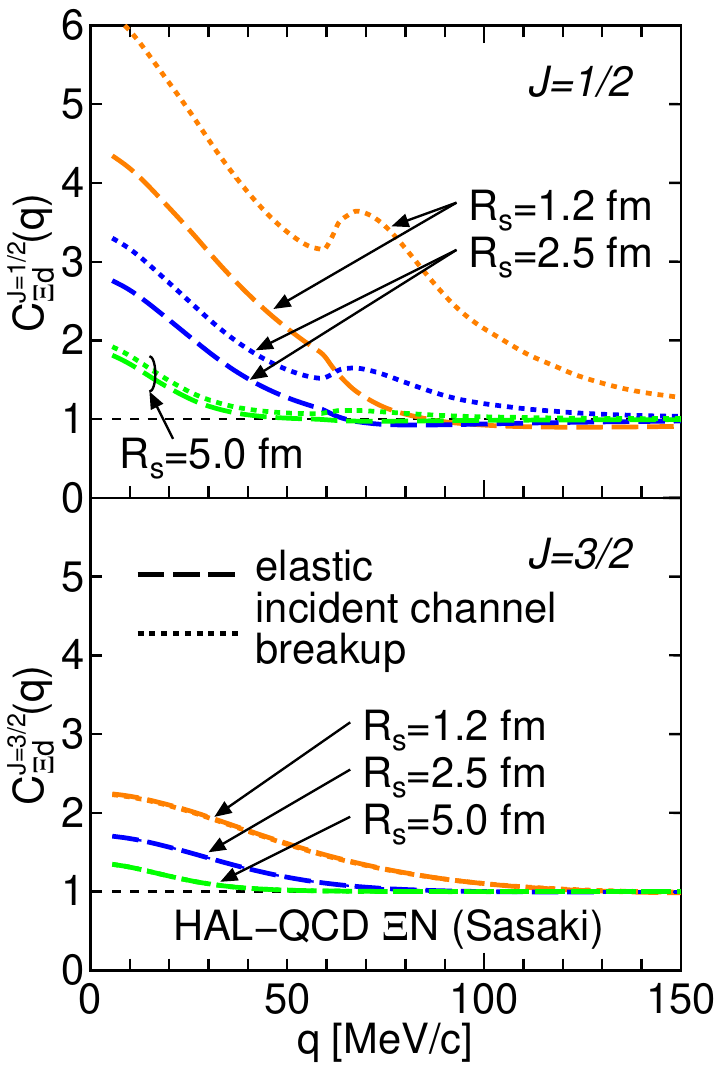}
\caption{Same as Fig. \ref{fig:incch}, but for the Sasaki $S=-2$ potential \cite{Sas20} and
the AV18 \cite{AV18} $NN$ interaction.
}
\label{fig:incsas}
\end{figure}

A cusp structure is observable in the $J=1/2$ elastic correlation functions
at $q\sim 60$, attributing to opening of the $^1$S$_0$ $NN$ decay at the deuteron
threshold, which is not present in the $\Lambda d$ correlation function \cite{KK25}.
In the presence of the deuteron breakup component, the strong coupling
to the $^1$S$_0$ $NN$ state exerts a substantial influence on the $\Xi d$ incident
channel wave function, particularly in the vicinity of the deuteron threshold.
No cusp structure is present in the $J=3/2$ correlation functions due to
the absence of the coupling to the $^1$S$_0$ $NN$ state. 

As demonstrated Figs. \ref{fig:incino} and \ref{fig:incsas}, the remaining two HAL-QCD
potentials yield qualitatively similar results. Fig. \ref{fig:incino} is for the Inoue potential,
and Fig. \ref{fig:incsas} is for the Sasaki potential.
Because these HAL-QCD potentials are parametrized using functions in entire
radial space, it is not appropriate to use the chiral N$^4$LO$^+$ NN interaction
with the HAL-QCD potentials. Therefore, we employ the AV18 interaction \cite{AV18}
for the $NN$ potential in these calculations.
The same vertical scale is employed in the upper and lower panels of each figure.
However, the scale varies depending on the figure. 

\begin{figure}[t]
\centering
 \includegraphics[width=0.45\textwidth,pagebox=cropbox,clip]{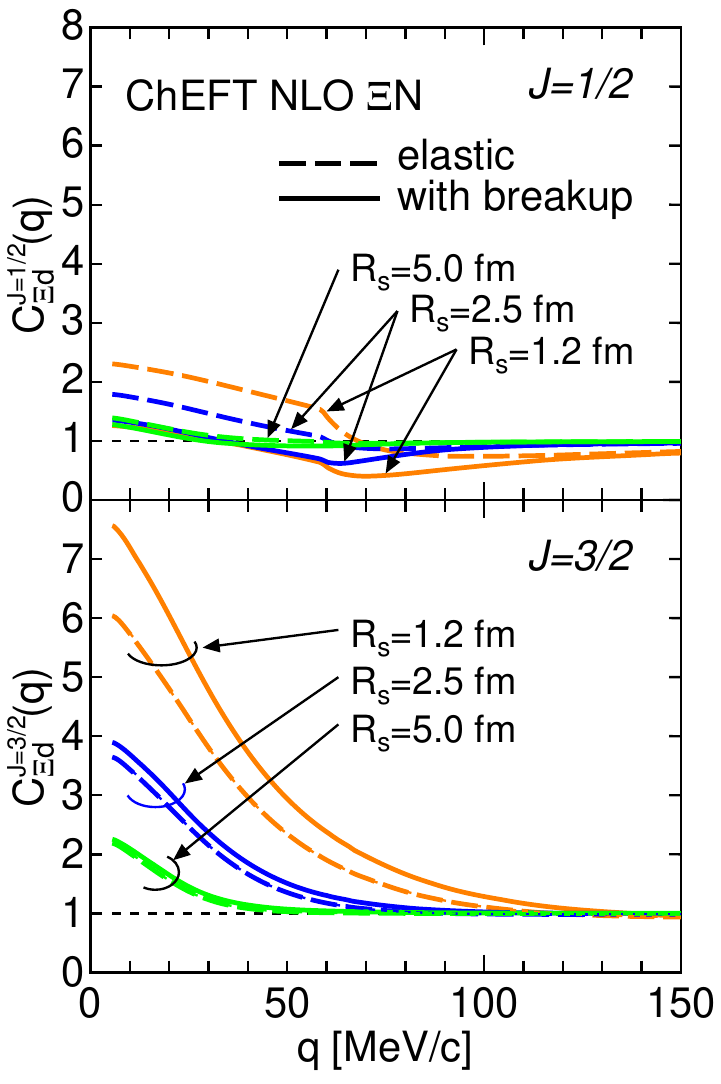}
\caption{$\Xi d$ momentum correlation functions in the $J=1/2$ and $3/2$ channels
for three choices of the source radius $R_s=1.2$, $2.5$, and $5.0$ fm, which are calculated
with the wave functions generated by the Faddeev calculations using the chiral NLO
$S=-2$ \cite{HAID19} and the chiral N$^4$LO$^+$ $NN$ \cite{RKE18} interactions.
The dashed curves are the results of the elastic wave functions.
The full three-body wave functions yield the solid curves.}
\label{fig:xidch}
\end{figure}

The differences of the absolute values of the correlation functions among three
$S=-2$ interactions reflect the properties of these spin-isospin properties illustrated
in the $\Xi N$ as well as $\Xi d$ phase shifts presented in Figs. \ref{fig:xnsph} and \ref{fig:xdph}. 

\begin{figure}[t]
\centering
 \includegraphics[width=0.45\textwidth,pagebox=cropbox,clip]{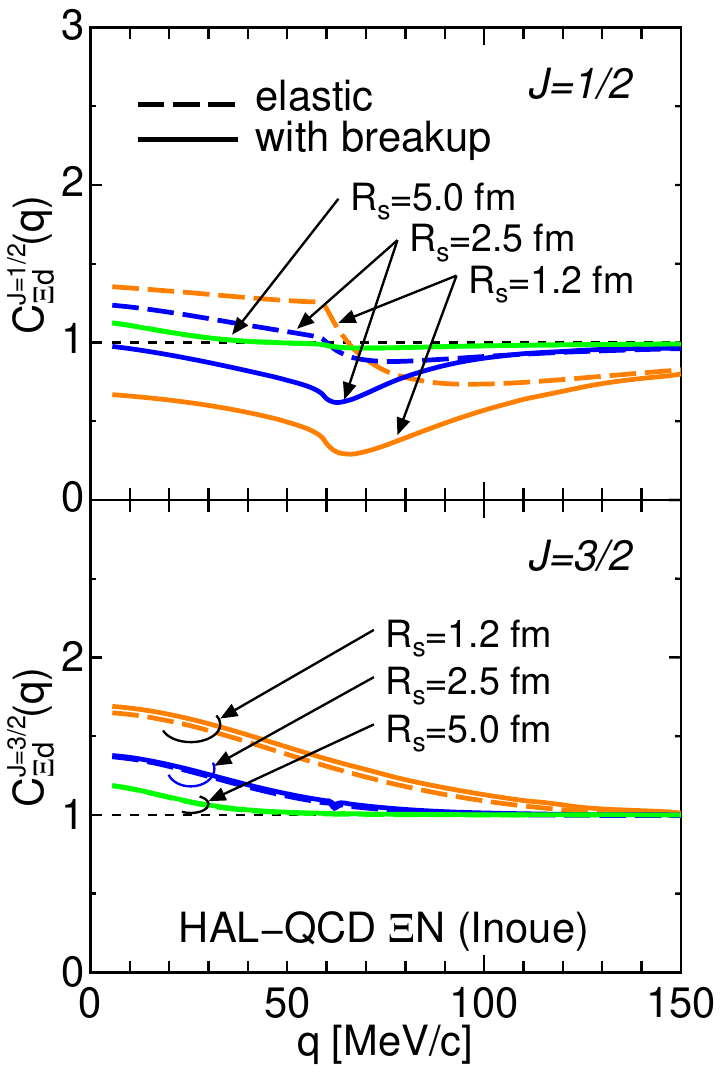}
\caption{Same as Fig. \ref{fig:xidch}, but for the Inoue $S=-2$ potential \cite{Ino19} and
the AV18 \cite{AV18} $NN$ interaction.
}
\label{fig:xidino}
\end{figure}

\begin{figure}[t]
\centering
 \includegraphics[width=0.45\textwidth,pagebox=cropbox,clip]{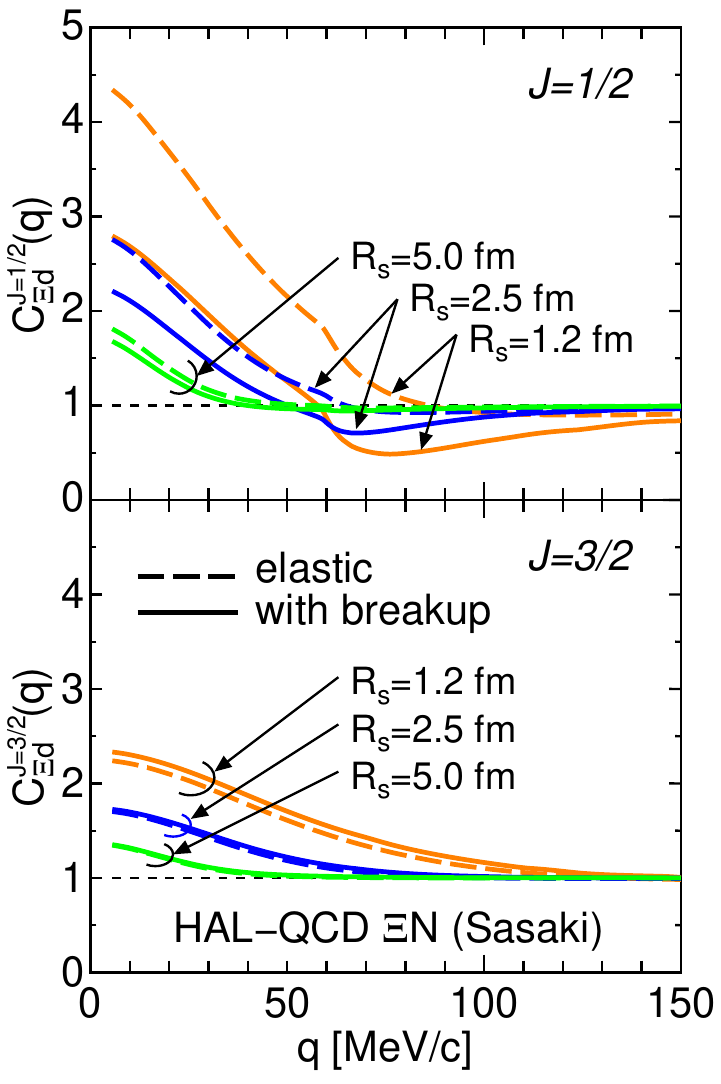}
\caption{Same as Fig. \ref{fig:xidch}, but for the Sasaki $S=-2$ potential \cite{Sas20} and
the AV18 \cite{AV18} $NN$ interaction.
}
\label{fig:xidsas}
\end{figure}

\subsection{Calculations with full three-body wave function}
The $\Xi d$ correlation functions calculated with the full three-body wave function, which
contains the component of the rearrangement channel, are presented in this subsection.
Results are shown in Figs. \ref{fig:xidch} - \ref{fig:xidsas} for the three $S=-2$
interactions. The vertical scale varies depending on the figure, but the scale in the upper
and lower panels is same in each figure. The dashed curves are the results of the elastic
wave function, which are equivalent to the dashed curves
in Figs. \ref{fig:incch} - \ref{fig:incsas}. 

The inclusion of the contribution from the rearrangement channel indicates
that the deuteron breakup process in the $J=3/2$ state results in an enhancement
of the correlation function in comparison to a negligible effect for the incident
channel wave function, particularly for the chiral NLO interaction.
In the $J=1/2$ state, the inclusion of the rearrangement channel causes
a substantial interference to decrease the correlation function, which becomes smaller
than the elastic one.

\begin{figure}[t]
\centering
 \includegraphics[width=0.45\textwidth,pagebox=cropbox,clip]{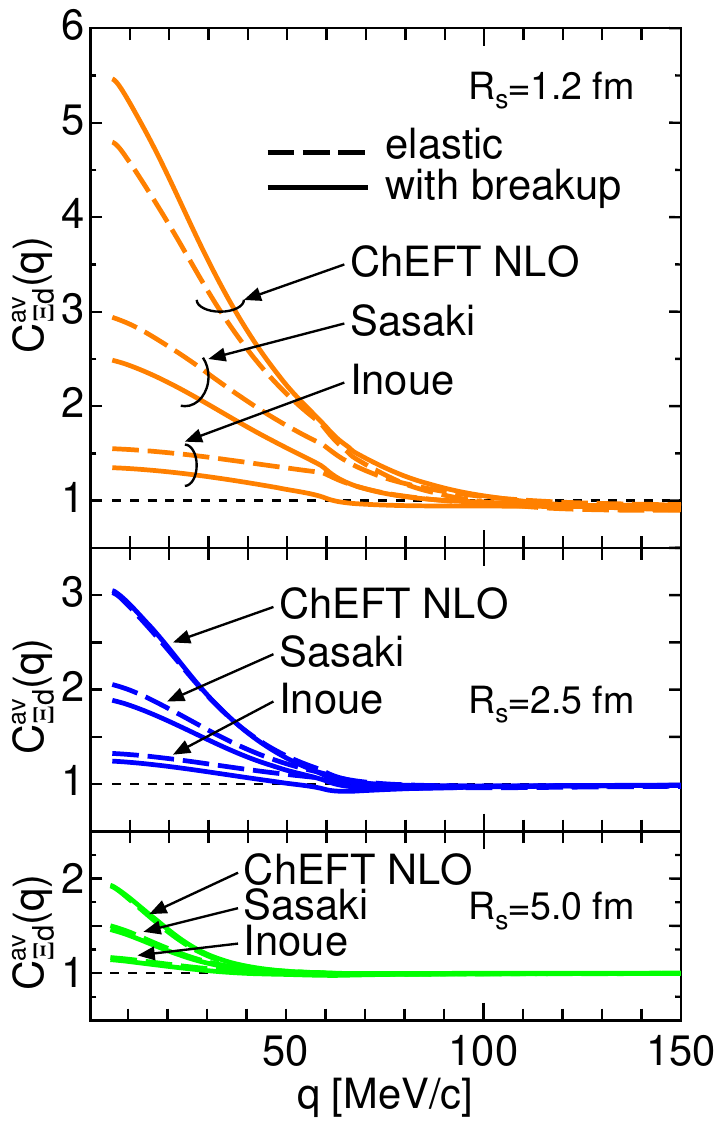}
\caption{Spin-averaged correlation functions, $C_{\Xi d}^{J-av}(q)=(C_{\Xi d}^{J=1/2}(q)
+2C_{\Xi d}^{J=3/2}(q))/3$,
for each source radius $R_s=1.2$, $2.5$ and $5.0$ fm with three $S=-2$ interactions:
the Chiral NLO \cite{HAID19}, Inoue \cite{Ino19}, and Sasaki \cite{Sas20}.}
\label{fig:avecor}
\end{figure}

Given that the experimental data is initially taken as a sum of the $J=1/2$ and $3/2$ states,
it is useful to show the total-spin averaged correlation function
$C^{av}(q)=(C^{J=1/2}+2C^{J=3/2})/3$, which is given in Fig. \ref{fig:avecor}. 
These results show a considerable dependence of the theoretical correlation function
on the three $\Xi N$ interactions employed. The depression below 1, as shown in the $J=1/2$
correlation functions in  Figs. \ref{fig:xidch} - \ref{fig:xidsas}, is masked in the average quantity. 

\section{Summary}
The understanding of the baryon-baryon interactions in a strangeness S=-2 sector
remains in its nascent stages, largely due to the paucity of precise scattering data.
Several experimental advances are reported in identifying $\Xi$ bound states in light
nuclei \cite{NAK15,HAY21,YO21,IC24}. However, it is difficult to infer the spin-isospin
structure of the bare two-body $\Xi N$ interactions, in which the coupling to the
$\Lambda$ and $\Sigma$ hyperons are involved, from the $\Xi$ binding energies
in nuclei. The recent development of measuring hyperon-proton and
hyperon-deuteron correlation functions in heavy-ion collision experiments has been
shown to provide reliable sources of information on baryon-baryon interactions.
These correlation functions are governed by the two-body and three-body wave
functions, which are determined by the interactions. While the theoretical analysis is
inherently model-dependent, it is important to pursue coherent descriptions of the
underlying baryon-baryon interactions.

In the present article, the $\Xi d$ $s$-wave scattering is investigated in a Faddeev
formulation, using the three representations of the $S=-2$ baryon-baryon interactions:
the chiral NLO $\Xi N$ interactions \cite{HAID19}, and two
different potentials by Inoue et al.  \cite{Ino19} and Sasaki et al.  \cite{Sas20}
that are based on the the HAL-QCD calculations. The $\Xi NN$ three-body
wave function, the asymptotic form of which is composed by the $\Xi$ incident
plane wave and the deuteron intrinsic bound-state wave function, is obtained
by the calculated Faddeev amplitudes. It is employed to calculated $\Xi d$
momentum correlation functions. 

In order to understand the characters of these $S=-2$ two-body interactions,
the $s$-wave phase shifts of the $\Xi N$ elastic scattering in each spin-isospin
channel are discussed in Sec. 2 by citing the results presented in Ref. \cite{KM21}.
Then the $\Xi d$ scattering phase shifts obtained by solving the Faddeev
equation for the $J=3/2$ and $J=1/2$ states are shown in Sec. 3.
The calculated phase shifts indicate that the $\Xi d$ interactions are moderately
attractive in both $J=3/2$ and $1/2$ channels, which is the result that
the $\Xi N$ interactions are attractive except in the $^{31}$S$_0$ state.
The differences of the absolute values of the phase shift among the employed
$\Xi N$ interactions indicate the varying characters in the strength of each
spin-isospin channel in these interactions. The behavior of the phase shifts
signifies that no $\Xi NN$ three-body bound state is expected, which is consistent
with the result of the Faddeev calculations for searching a possible bound state
in Ref. \cite{MK21}, as far as the $\Xi N$ interactions employed are concerned.

The $\Xi d$ three-body wave function in coordinate space is derived from the
calculated Faddeev amplitudes in momentum space. In addition to the full
three-body wave function, it is possible to consider the wave function
in the incident channel to elucidate deuteron breakup effects in the relevant
channel, though it is not directly related to the observed correlation function.

The $\Xi d$ momentum correlation functions employing these wave functions
for the theoretical expression explained in Sec. IV are presented in Sec. V.
First, the effects of the deuteron breakup in the incident channel are
demonstrated. Then the results calculated by the full three-body wave
function are shown. The cusp at the opening of the $^1$S$_0$ NN state
is evident in the $J=3/2$ channel. Besides, the effects of the deuteron breakup
are found to be sizable.
As stated in Introduction, it is important to note that the present calculations do not
consider the Coulomb force or the mass difference between $\Xi^-$ and $\Xi^0$.
Nevertheless, the results elucidate the quantitative differences of the calculated
$\Xi d$ momentum correlation functions both in the $J=1/2$ and $J=3/2$ channels,
depending on the $\Xi N$ interactions employed. The prospective experimental data
on the $\Xi d$ momentum correlation function could contribute to a better description
of the $S=-2$ interactions, which serves to yield coherent description of the $\Xi$
hyperons in the nuclear medium.

\bigskip
{\textit{Acknowledgements.}}
We are grateful to K. Miyagawa for his valuable discussions and comments on this study.
This work is supported by Japan Society for the Promotion of Science (JSPS) KAKENHI
Grants No. JP22K03597, No. JP24K07019, and No. JP25K07301.

\appendix
\section{Faddeev equations}
The essential equations and expressions for calculating $\Xi d$ phase shifts and
momentum correlation functions are outlined in this Appendix for completeness. 

\begin{figure}[b]
\centering
\includegraphics[width=0.3\textwidth]{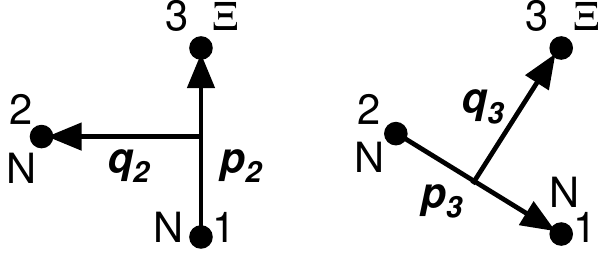}
\caption{Two sets of Jacobi momenta. The state designated by ($\bp_3,\bq_3$) is
referred to as an incident channel.}
\label{fig:jacobi}
\end{figure}

The Faddeev equations for the $\Xi d$ scattering are formulated as simultaneous equations
for the two operators $T_2$ and $T_3$, as follows \cite{KK24,GL96}:
\begin{align}
 \langle p_3 q_3 \alpha_3|T_3|\phi\rangle =& \langle p_3 q_3 \alpha_3|t_3 G_0 (1-P_{12})T_2
 |\phi\rangle, \label{eq:pe1}\\
 \langle p_2 q_2 \alpha_2|T_2|\phi\rangle =& \langle p_2 q_2 \alpha_2|t_2 + t_2 G_0 T_3
 - t_2 P_{12} G_0  T_2 |\phi\rangle,
\label{eq:pe2}
\end{align}
where $|p_i q_i \alpha_i\rangle$ ($i=2,\;3$) is a partial wave projected state and
$|\phi\rangle=|\phi_d, q_0\rangle$ is an incident wave function which is a product
of a free deuteron wave function $\phi_d$ times a $\Xi$-deuteron plane wave with
its momentum $q_0$. $G_0=[E-H_0]^{-1}$ is a free three-body Green function
for the total energy $E$ with a free Hamiltonian $H_0$, $P_{12}$ is a nucleon-nucleon
exchange operator, and $t_i$ is a pertinent two-body $t$-matrix. Two sets of
the Jacobi momenta $(\bp_i,\bq_i)$ with $i=2,3$ are defined as in Fig. \ref{fig:jacobi}.
The indicator $\alpha_i$ is used to specify the partial-wave channel. The $\Lambda$ and
$\Sigma$ hyperons are not  included in the basis states, although the couplings of the
$\Xi N$ system to the $\Lambda \Lambda$, $\Lambda \Sigma$ and $\Sigma\Sigma$
state coupling are taken care of in evaluating the two-body $t_2$ matrix.

The spatial coordinates corresponding to the Jacobi momenta $(\bp_3,\bq_3)$ are denoted by
$(\br_{np},\br_3)$. The three-body wave functions in the configuration space are derived
from the obtained matrix elements of the operators $T_3$ and $T_2$ in momentum space.

The incident channel wave function in the configuration space
is constructed from the Faddeev amplitude as:
\begin{align}
   \psi(\br_{np},\br_{3})= 
  \langle \br_{np},\br_3|\phi\rangle +\langle \br_{np},\br_3|G_3 (1-P_{12})T_2|\phi\rangle,
\label{eq:chwf}
\end{align}
where $G_3$ is a Green function of the incident channel Hamiltonian $H_0+V_{12}$:
\begin{equation}
 G_3= [E-H_0-V_{12}+i\epsilon]^{-1},
\end{equation}
where $V_{12}$ is an $NN$ interaction. The spectral representation of the incident
channel Green function $G_3$ consists of a bound-state part and a scattering part
for the partial wave in which the deuteron is involved. The bound-state part
represents a $\Xi d$ elastic wave function.

The full three-body wave function is given by the Faddeev amplitudes as follows \cite{GL96}.
\begin{align}
 & \psi(\br_{np},\br_{3})= \langle \br_{np},\br_3|G_0 \{G_0^{-1}+2T_2+T_3\}|\phi\rangle.
\label{eq:fwf}
\end{align}
This wave function contains the deuteron breakup effects in both the initial and rearrangement
channels.

Further details such as the treatment of the deuteron pole and the procedure of numerical
calculations of Eqs. (A4) and (A5) are to be referred to Ref. \cite{KK25}.

\end{document}